# System Identification of a Multi-timescale Adaptive Threshold Neuronal Model[1]

**Amirhossein Jabalameli**[†]

**Aman Behal**[†, ‡]

[†] Department of Electrical Engineering and Computer Science, University of Central Florida, Orlando, FL 32816.

[‡] NanoScience Technology Center, University of Central Florida, Orlando, FL 32826.

## Abstract

[1] An abridged version (Jabalameli and Behal, 2015) of this paper was presented at the 2015 International Conference on Computational Advances in Bio and Medical Sciences. The study was supported by Award # R15NS062402 from NINDS. The content is solely the responsibility of the authors and does not necessarily represent the official views of the NINDS or the NIH.

In this paper, the parameter estimation problem for a multi-timescale adaptive threshold (MAT) neuronal model is investigated. By manipulating the system dynamics, which comprise of a non-resetting leaky integrator coupled with an adaptive threshold, the threshold voltage can be obtained as a realizable model that is linear in the unknown parameters. This linearly parametrized realizable model is then utilized inside a prediction error based framework to identify the threshold parameters with the purpose of predicting single neuron precise firing times. The iterative linear least squares estimation scheme is evaluated using both synthetic data obtained from an exact model as well as experimental data obtained from *in vitro* rat somatosensory cortical neurons. Results show the ability of this approach to fit the MAT model to different types of fluctuating reference data. The performance of the proposed approach is seen to be superior when comparing with existing identification approaches used by the neuronal community.

# 1   Introduction

As large-scale detailed network modeling projects are appearing in the computational neuroscience area, it becomes essential to construct easily identifiable lower dimensional models of single neurons. While lower dimensional models allow for construction of large-scale computational neuronal networks that can be simulated with ease, identifiability of the underlying neuronal models is necessary for the neuronal network to be able to mimic the computational properties of the biological structure being modeled *in silico*. Fortunately, a wide variety of single neuron models are available in literature. These models can broadly be categorized into two main groups: detailed biophysical models and simple phenomenological models, *e.g.*, (Izhikevich, 2004; Jolivet et al., 2008b). Detailed biophysical Hodgkin-Huxley (Hodgkin and Huxley, 1952) type neuron models can accurately reproduce most behaviors of neurons, however, their complex dynamics and a high-dimensional parameter space make them an impractical choice as building blocks for large scale neuronal networks (Zhi et al., 2012). In spiking neuron models, since isolated spikes of a given neuron are similar, the shape of the action potential does not represent any information. Instead, it is the spike train, as a sequence of spikes, which is important (Gerstner et al., 2014). Due to this reason and high computational costs, simple phenomenological models such as leaky integrate-and-fire models have been proposed (Brunel and van Rossum, 2007; Stein, 1965) and developed to study the dynamics of neural networks (Gerstner and Kistler, 2002; Izhikevich, 2004). Recently, substantial efforts have been put in the expansion of leaky integrate and fire models for fitting of such models to data in order to reproduce quantitative features (Prinz et al., 2003; Huys et al., 2006). Several methods have been proposed that can accurately predict the timing of spikes, *e.g.*, (Jolivet and Gerstner, 2004; Kobayashi et al., 2009; Kobayashi and Shinomoto, 2007; Clopath et al., 2007).

Identification of model parameters can be performed by several methods. Although hand tuning of parameters may yield reasonable results, this process is labor intensive and impractical. In our previous work (Zhi et al., 2012), the versatile quadratic model proposed by Izhikevich (Izhikevich, 2003) was utilized to automatically identify experimentally obtained neuronal firing data. However, as noted in (Chen et al., 2011, 2016), that model cannot be utilized for identification because it is unable to quantitatively represent the upstroke of the spike unless it is assumed that the model parameters are voltage-dependent (Izhikevich, 2007). In Jabalameli (2015), a two-stage linear parameter estimation strategy was explored based on a five parameter subthreshold model coupled with a voltage dependent threshold model but the estimation results were seen to be unsatisfactory.

Another proposed model, *viz.*, the Multi-timescale Adaptive Threshold model (MAT) (Kobayashi et al., 2009) shows great performance for both stationary and non-stationary fluctuating currents to replicate spike trains of experimental data (Yamauchi et al., 2011). While the threshold is dependent on five independent parameters, the authors *a priori* fix two time constant related model parameters and identify the other three parameters by maximizing a non-convex performance metric encoding for the coincidence of spikes between the model predictions and the experimental data. Thus, a systematic technique is still needed to automatically identify all five parameters so that the threshold model firing pattern is consistent with that of the experimental data. Furthermore, a convex cost function is needed in order to guarantee parameter convergence. The advantage of such a technique is that it can be utilized in a fully automated system that can identify the underlying neuronal models that are needed to design and implement a realistic biological computational structure. This work focuses on the development of the aforementioned automated identification technique. Specifically, we manipulate the original threshold equation (which is nonlinearly dependent on its parameters) into a linear-in-the-parameters model and then proceed to estimate the parameters in order to minimize in a least squares sense the error between the subthreshold voltage and the threshold estimate at the experimentally observed spiking times. To ensure meaningfulness of the values of the obtained parameter estimates, convex constraints are generated and imposed on the optimization. Results show that the proposed scheme outperforms existing strategies in terms of reproducing spike locations. Another novelty of the proposed method is that unlike many other methods that require the input and the reference membrane voltage of the neuron to tune their models, this method only needs the input current and the spike locations to fit the model to the reference data.

The remainder of this paper is organized as follows. The basics of the multi-timescale adaptive threshold (MAT) model are introduced in Section 2. Next, in Section 3, we pursue the algebraic manipulations needed to arrive at the proposed scheme to identify the MAT model parameters. Section 4 describes the steps of implementation. In Section 5, results are provided on model identification from different types of reference data followed by comparisons with results from existing approaches. Pertinent conclusions are drawn in Section 6.

## 2 The MAT Model

The Multi-timescale Adaptive Threshold (MAT) model (Kobayashi et al., 2009) was proposed by Kobayashi *et al.* for the purpose of predicting the timing of output spikes of neurons. The MAT model can be described by a subthreshold voltage $V$ and a multi-timescale adaptive threshold $f$. While it is possible in general to have an arbitrary number of timescales, the analysis in this paper will be limited to two timescales as similarly done in (Kobayashi et al., 2009). The subthreshold voltage can be obtained by the leaky integrator which is given by a first order differential equation (1)

$$\tau_m \frac{dV}{dt} = -V(t) + RI(t) \tag{1}$$

where $V(t)$ denotes a membrane potential, $I(t)$ is the injected input current, while $\tau_m$ and $R$ are parameters that describe leaky time constant and input resistance, respectively. While Equation (1) is the foundation of Generalized Linear Models and Spike Response Model (Gerstner et al., 2014), however, in the MAT model, the variable $V(t)$ is not reset after reaching a constant or time/state dependent threshold. Instead, at the spiking instants defined by the intersection of $V$ and $f$, the threshold variable $f$ resets to a different value in the manner shown below

$$f(t) = \alpha_1 \sum^{t_k} \exp(-k_1(t-t_k)) + \alpha_2 \sum^{t_k} \exp(-k_2(t-t_k)) + \omega \tag{2}$$

where $t_k$ is the $k^{th}$ spike time, $k_1$ and $k_2$ denote inverses of the two threshold time constants, $\alpha_1$ and $\alpha_2$ denote the increments of the threshold at the spike instants, while $\omega$ is the threshold resting value. Furthermore, an absolute refractory period $\tau_R$ is defined to prevent consecutive firing; consequently, within $\tau_R$ period after a spike, the model cannot fire more spikes even if the subthreshold voltage is above the threshold. As seen in Figure (1), spikes (represented by arrows) are assumed to be generated whenever the subthreshold voltage reaches the threshold voltage from below. The moment of crossing is the so-called firing time at which instant the threshold voltage increases a certain amount and then starts decaying exponentially to its resting value.

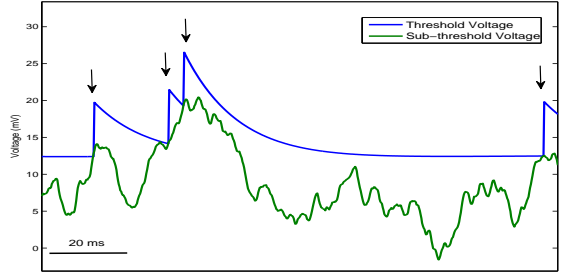

Figure 1: Dynamics of the MAT model. When the threshold voltage (blue) intersects the subthreshold voltage (green), a spike is generated and the threshold jumps.

# 3 The Proposed Estimation Technique

Pursuant to the development in the previous section, the MAT model can be completely characterized by 7 free parameters; where $\{\tau_m, R\}$ are the leaky integrator parameters and $\{\alpha_1, \alpha_2, k_1, k_2, \omega\}$ are the spike threshold parameters. The resistance $R$ does not affect the spike time prediction and only scales the subthreshold voltage (Kobayashi et al., 2009); furthermore, a common membrane time constant ($\tau_m$) is extracted from the data and preselected for all simulations. Therefore, we focus our work to estimate the threshold parameters and we assert that $\theta_0 \triangleq [\alpha_1, \alpha_2, k_1, k_2, \omega]^T$ completely describes the model. In what follows, we develop an automatic method for estimating the MAT model parameters.

## 3.1 Linear Parametrization

While the static threshold representation of (2) is nonlinear with respect to the parameters $k_1$ and $k_2$, a dynamic representation can be developed to acquire a linearly parameterized model. By taking Laplace transform of (2) and rearranging the terms, one can

obtain

$$(s^2)F(s) = -[(k_1 + k_2)s + k_1k_2]F(s) + [(\alpha_1 + \alpha_2)s \\
+ (\alpha_1 k_2 + \alpha_2 k_1)]\sum^{t_k} \exp(-t_k s) \qquad (3) \\
+ [ws^2 + w(k_1+k_2)s + wk_1k_2]\frac{1}{s}$$

where $s$ is the Laplace variable and $F(s)$ represents the Laplace transform of $f(t)$. A second order low pass filter

$$\frac{1}{A} = \frac{1}{s^2 + \beta_1 s + \beta_0} \qquad (4)$$

is employed in order to eliminate the model dependency on derivatives of the measurable signals (Zhi et al., 2012). The following can be obtained by applying the filter to both sides of (3):

$$F(s) = (\frac{\beta_1 s + \beta_0}{A})F(s) - (k_1+k_2)\frac{s}{A}F(s) - k_1 k_2 \frac{1}{A}F(s) \\
+ (\alpha_1+\alpha_2)\frac{s}{A}\sum^{t_k}\exp(-t_k s) + (\alpha_1 k_2 + \alpha_2 k_1) \\
\frac{1}{A}\sum^{t_k}\exp(-t_k s) + wk_1 k_2 \frac{1}{A}\frac{1}{s} + \\
\overbrace{[w\frac{s^2}{A}\frac{1}{2} + w(k_1+k_2)\frac{s}{A}\frac{1}{s}]}  \qquad (5)$$

Since the last row of (5) vanishes beyond an initial transient, we can neglect it in the subsequent calculations to reduce the dimension of the parameter vector. Ultimately, the linearly parameterized (LP) model is described as follows

$$f(t) = \Psi^T(f, t, t_k)\theta + \Phi(f, t) \qquad (6)$$

where $\Phi(f, t)$ is a signal that is independent with respect to the model parameters and $\Psi(f, t, t_k) \in \mathbb{R}^5$ is a regression vector. These signals are defined as follows

$$\Phi(f, t) \triangleq L^{-1}\left\{(\frac{\beta_1 s + \beta_0}{A})F(s)\right\} \qquad (7)$$

$$\Psi(f,t,t_k) \triangleq L^{-1}[\tfrac{s}{A}F(s), \tfrac{1}{A}F(s), \tfrac{s}{A}\sum^{t_k}\exp(-t_k s), \tag{8}$$
$$\tfrac{1}{A}\sum^{t_k}\exp(-t_k s), \tfrac{1}{sA}]^T$$

while $\theta \in \mathbb{R}^5$ is an unknown auxiliary parameter vector that is a nonlinear function of $\theta_0$ and defined as follows

$$\theta = [-(k_1+k_2), -k_1 k_2, \alpha_1 + \alpha_2, \alpha_1 k_2 + \alpha_2 k_1, w k_1 k_2]^T. \tag{9}$$

## 3.2 Algorithm Development

According to the MAT model process, firing happens whenever the subthreshold voltage reaches the threshold voltage. In other words, $f(t)$ and $V(t)$ are equal at the spike instants. Since the main objective in this work is to fit the MAT model to the reference data in order to obtain a predictive model for the location of spike times, we define a prediction error variable as follows for each spike moment

$$e_{t_k} \triangleq \hat{f}(t_k) - V(t_k) \tag{10}$$

where $\hat{f}(t)$ denotes an estimate of $f(t)$. A natural cost function based on the prediction error for all spike instants can be defined as

$$J \triangleq \sum^{t_k} e_{t_k}^2$$

and developed as follows

$$J = \sum^{t_k}(\hat{f}(t) - V(t))^2 = \sum^{t_k}(\underbrace{\Psi(f,t,t_k)\hat{\theta} + \Phi(f,t)}_{\hat{f}(t)} - V(t))^2 \tag{11}$$

based on the LP model derived in (6). Here, $\hat{\theta}$ denotes an estimate for $\theta$. Therefore, the objective is to find the MAT model parameters that minimize the prediction error

$$\hat{\theta} = \arg\min\left\{J = \sum^{t_k}(\hat{f}(t) - V(t))^2\right\} \tag{12}$$

However, certain constraints need to be introduced so that the parameter estimates $\hat{\theta}$ converge to physically meaningful values when mapped back to the actual parameter space $\theta_0$. The actual model parameters can be obtained in closed form as follows

$$k_1 = \max(\frac{-\theta_1 \pm \sqrt{\theta_1^2 + 4\theta_2}}{2}) \qquad (13)$$

$$k_2 = \min(\frac{-\theta_1 \pm \sqrt{\theta_1^2 + 4\theta_2}}{2}) \qquad (14)$$

$$\alpha_1 = \frac{\theta_4 - k_1\theta_3}{k_2 - k_1} \ , \ \alpha_2 = \theta_3 - \alpha_1 \ , \ w = \frac{-\theta_5}{\theta_2} \qquad (15)$$

where $\theta_i$ denotes the $i^{th}$ component of $\theta$. While there are no restrictions on $\alpha_1, \alpha_2$, and $w$ other than that they are real (which is trivially enforced), $k_1$ and $k_2$, being time constants, are needed by the model to be positive and real. Thus, the set of equations above suggest the inequalities

$$\theta_1, \theta_2 < 0 \text{ and } \theta_1^2 + 4\theta_2 \geq 0 \qquad (16)$$

to ensure the real positiveness of $k_1$ and $k_2$. As seen in Figure 2, these inequalities define a non-convex region. Since $k_1$ and $k_2$ denote different timescales, a separation based on available data and studies suggests the choice $20 \leq k_1 \leq 500$, $2 \leq k_2 \leq 40$ (the unit is $1/\sec$). Based on this feasible range, the following set of convex constraints can be obtained

$$-540 \leq \theta_1 \leq -22 \ , \ -2 \times 10^4 \leq \theta_2 \leq -40$$
$$38.5\theta_1 - \theta_2 \leq -1482 \qquad (17)$$
$$-1.7\theta_1 + \theta_2 \leq 0$$

which is a triangular area demarcated in Figure 2 by the dashed blue and green lines and the vertical solid black line. The above set of constraints does not preclude intersection of $V$ and $f$ between the experimentally observed spiking instants as shown in Figure 3. However, it is impractical to check for and enforce this condition at all times. Instead, we endeavor to enforce this constraint practically by observing the maximum value of $V(t)|_{t=t_m}$ between each pair of spikes and enforcing the following convex constraint between each pair of spikes:

$$f(t_m) - V(t_m) = \Psi^T(t_m)\theta + \Phi(t_m) - V(t_m) > 0. \qquad (18)$$

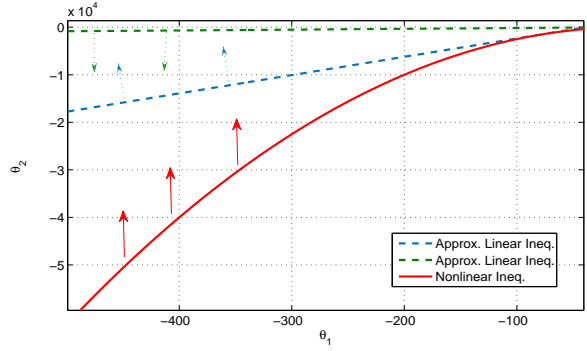

Figure 2: Feasible Region for Solution

This constraint essentially states that the subthreshold voltage at its peak between any pair of experimentally observed spikes is not allowed to cross over the threshold. Of course, this set of constraints does not avoid intersection altogether since the threshold is time varying but it is an easily implementable set of constraints that serves well to improve the model efficacy by reducing false positives as will be seen in the sequel.

## 4 Implementation Procedure

Although no iterative procedure is apparent at first glance when solving the constrained least squares problem defined by (12), (17) and (18), the unavailability of the signal

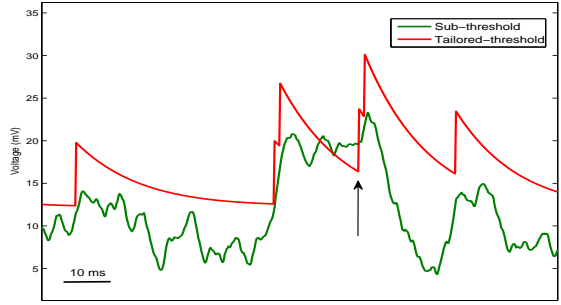

Figure 3: Per the objective function, the method only minimizes the error at the spike times without considering possible intersections of subthreshold (green) and threshold voltage (red) between spike times.

$f$ on the RHS of (11) forces us instead to replace $f$ with $\hat{f}\left(\hat{\theta}_{k-1}\right)$ at the $k^{th}$ iteration and apply the algorithm iteratively until the parameters converge. The implementation proceeds according to the following steps:

**Step 1:** Generate subthreshold voltage $V$: Since the objective function (12) requires sub-threshold voltage $V$, we solve (1) by assuming $R = 50M\Omega$ and $\tau_m = 5\ ms$. Following (Jolivet et al., 2008b), the excitation is performed with current generated from an Ornstein–Uhlenbeck process as follows

$$I(t+dt) = I(t) - \frac{I(t)}{\tau_I}dt + m_I dt + s_I \zeta(t)\sqrt{dt} \quad (19)$$

where $m_I$ and $s_I$ are parameters and $\zeta(t)$ is a zero-mean, unit variance Gaussian random variable. Similar to (Jolivet et al., 2008b), the process is generated and injected at a rate of 5 kHz and the correlation length $\tau_I$ is $1\ ms$. The resulting current $I(t)$ has a stationary Gaussian distribution with mean $\mu_I = m_I \tau_I$ and variance $\sigma_I^2 = s_I^2 \tau_I/2$.

**Step 2:** Build signals $\Psi(\cdot)$ and $\Phi(\cdot)$ by using (7) and (8): Since both are dependent on $f$ which is not available for measurement, we begin with an initial guess $\hat{\theta}_0$ for $\theta_0$ and build $\hat{f} \triangleq f(\hat{\theta}_0, t, t_k)$ based on which we build $\hat{\Psi}(\cdot) \triangleq \Psi(\hat{f}, t, t_k)$ and $\hat{\Phi}(\cdot) \triangleq \Phi(\hat{f}, t)$. During estimation computations, $f(t)$ does not reset at the times where the threshold crosses the subthreshold, instead, the generated $f(t)$ fires at the reference (experimentally obtained) spike times, $t_k$.

**Step 3:** Minimize the objective function (11) to obtain $\hat{\theta}$ subjects to the set of constraints (17) and (18). The loop error is defined as follows:

$$e(\hat{f}) \triangleq \sum^{t_k}(\Psi(\hat{f}, t, t_k)\hat{\theta} + \Phi(\hat{f}, t) - V(t))^2 \qquad (20)$$

where $e(\hat{f})$ denotes the error of $\hat{f}$ that is generated by $\hat{\theta}$ at the end of the loop. Since the error function is a quadratic function of variables which is subject to linear constraints on those variables, the optimization problem (12) at each step is formulated as a Quadratic Programming problem.

**Step 4:** Solve $\hat{\theta}$ to obtain $\hat{\theta}_0$ according to (13)-(15).

**Step 5:** Finally, $\hat{\theta}_0$ is updated at Step 2 with new parameters and the procedure is repeated until all parameters converge to constant values.

## 4.1 Evaluation of Prediction

The error function defined in (11) is useful for estimating parameters but it cannot be an evaluation criterion since the main aim of this work is to predict the spike train produced by the neurons. While the firing rate of a spike train provides helpful information, yet, the evaluation of similarity between two spike trains is needed to capture local artifacts. Several measures exist for comparing the spike train predicted by the model and the spike train generated by the reference data. A popular index known as the *coincidence factor* has been proposed in (Jolivet et al., 2008b). This coincidence factor $\Gamma$ measures both the similarity and dissimilarity of two spike train by considering the spiking rate and coincident spikes. $\Gamma$ is calculated as follows (Jolivet et al., 2008b)

$$\Gamma = \frac{N_{Coinc} - <N_{Coinc}>}{N_{Data} + N_{Model}} \times \frac{2}{1 - 2\upsilon\Delta} \quad (21)$$

where $N_{Data}$ is the number of spikes in the reference spike train, $N_{Model}$ is the number of spikes in the predicted spike train, $N_{Coinc}$ is the number of coincidences with precision $\Delta$ between the two spike trains, and $<N_{coinc}>$ is the expected number of coincidences generated by a homogeneous Poisson process with the same rate $\upsilon$ as the spike train of the model. The factor $2/(1 - 2\upsilon\Delta)$ normalizes $\Gamma$ to a maximum value of 1 which is reached if and only if the spike train of the model reproduces exactly the reference spike train. Hence, after identifying the model parameters, we calculate the value of the Coincidence Factor to evaluate the predicted spike times.

# 5 Results and Discussion

In this section, the parameters of the MAT model are identified by the proposed method to match three types of reference data: (a) data from the MAT model, (b) noisy version of data from the MAT model, and (c) experimental data. The synthetic datasets are used to test validity and robustness of our approach. For our experimental data, we utilized a standard dataset from a Quantitative Neuron Modeling competition (Jolivet et al., 2008b) which includes the excitation input and single-electrode data recorded from a cortical pyramidal neuron in slices of rat barrel cortex. The details of the experimental protocol are available in (Rauch et al., 2003) and (La Camera et al., 2006). The injected input current is generated based on (19) and stimulation is done with currents of differ-ent means and fluctuation amplitudes. Additionally, for computing Coincidence Factor, the value of $\Delta$ is set to $2\ ms$, since it is in the same range as the accuracy of measuring synaptic rise times in the soma of cortical pyramidal neurons (Jolivet et al., 2008b).

## 5.1 Results from MAT Reference Data

First, to show the ability of the proposed method to identify the model parameters, a reference train of spikes is produced by the exact MAT model using a known set of parameters shown in Table (1). The subthreshold voltage $V$ is generated via the method described in Section 4 and is assumed to be noiseless. An initial guess for $\theta_0$ (reasonably far away from the actual parameter values) is chosen as follows

$$\alpha_1 = 10 \quad \alpha_2 = 5 \quad k_1 = 50 \quad k_2 = 8 \quad \omega = 13$$

Following the procedure in Section 4, the parameters are seen to converge to final values as shown in Table (1). It is seen that the estimated values are very close to the actual model parameters. Figures (4) and (5) represent the convergence of the objective function value and the estimated parameters along the estimation loops. To illustrate the evolution of the threshold function, Figure (6) shows the generated $f(t)$ in a few sample process loops.

Table 1: Identified and Actual Parameters for the MAT reference data.

| $\theta_0$ | $\alpha_1$ | $\alpha_2$ | $k_1$ | $k_2$ | $\omega$ |
|---|---|---|---|---|---|
| Identified Val. | 3.93 | 0.48 | 98.39 | 4.71 | 15.13 |
| Actual Val. | 4 | 0.5 | 100 | 5 | 15 |

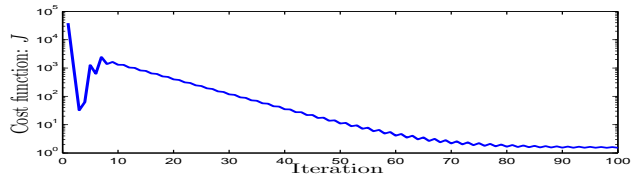

Figure 4: Objective function error. The vertical axis is log scale and indicates the value of the error function along the estimation loops

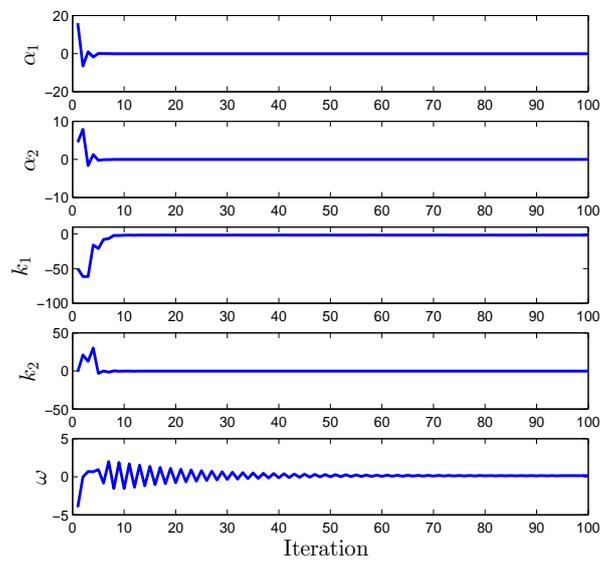

Figure 5: Evolution of identified parameters $\alpha_1, \alpha_2, k_1, k_2$ and $\omega$ using noise-free synthetic data

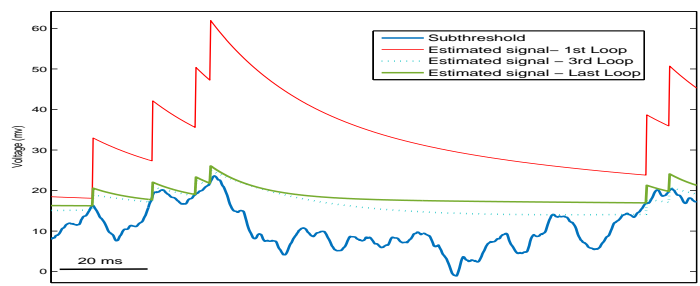

Figure 6: Convergence of estimated threshold during the loops

## 5.2 Results from Noisy MAT Reference Data

In a neural recording experiment *in vivo,* the input current received by the neuron is divided into two components, a deterministic one and a stochastic one (Gerstner et al., 2014). The deterministic part does not vary during the trials with the same stimulus and the stochastic part represents all the remaining inputs which change during the trials. Therefore, to consider noise in the biological system, the stochastic component of the input is treated as noise which is added to the right hand side of the subthreshold voltage dynamics. Specifically, we modified the subthreshold voltage by adding Gaussian White Noise (GWN) to it. Results from a simulation with $SNR = 35dB$ can be seen in Table (2) and Figures (7)-(9). Table (2) shows that the estimated values are very close to the actual model parameters. A comparison with Table (1) shows that the estimates are quite robust with respect to noise at this level of SNR. Similar to Section 5.1, Figures (7), (8), and (9) show the convergence of, respectively, the objective function value, the estimated parameters, and the threshold function along the estimation loops.

Table 2: Identified and Actual Parameters for noisy ($SNR = 35$) MAT reference data.

| $\theta_0$ | $\alpha_1$ | $\alpha_2$ | $k_1$ | $k_2$ | $\omega$ |
|---|---|---|---|---|---|
| Identified Val. | 4.05 | 0.48 | 100.30 | 5.49 | 15.34 |
| Actual Val. | 4 | 0.5 | 100 | 5 | 15 |

We performed simulations for different SNR values to measure robustness of the proposed method against noise. Our simulations show that the randomness of the generated noise causes the parameters to not converge during some trials. Table (3) describes the obtained results for 10 trials of performed simulations. It is seen that the rate of non-convergence and the average error per spike over the converged trials ($\bar{e}$) decreases as SNR becomes better.

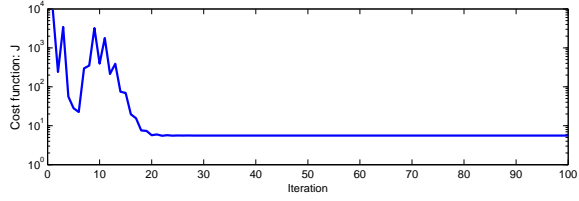

Figure 7: Cost Function Evolution with Iteration using Noisy Synthetic Data

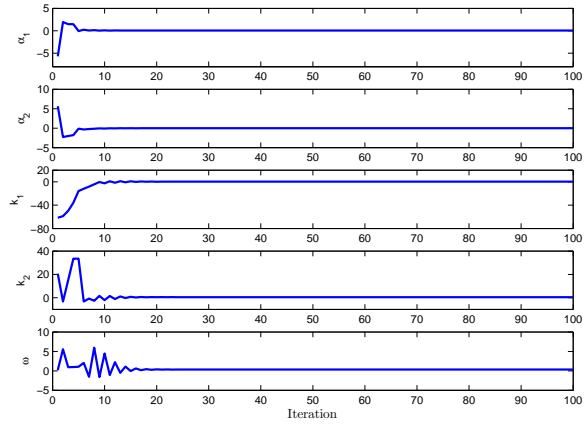

Figure 8: Evolution of Parameter Errors with Noisy Synthetic Data

Table 3: Estimation results for noisy MAT model reference data

|             | $SNR = 40$ | $SNR = 35$ | $SNR = 30$ |
|-------------|------------|------------|------------|
| Convergence | 8/10       | 7/10       | 5/10       |
| $\bar{e}$   | 0.0273     | 0.0502     | 0.1669     |

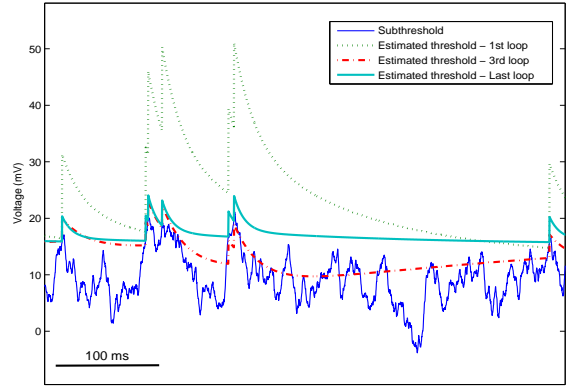

Figure 9: Convergence of Threshold Trace during Iterative Process

## 5.3 Results from Experimental Data

After confirming the ability of the proposed method to identify the exact model reference data, the estimation procedure is applied to *in vitro* experimental data from the Single Neuron Competition (Challenge A, 2007). Table (4) presents 3 different parameter estimation results obtained by applying the proposed algorithm to 3 experimental datasets.

Table 4: Identified Parameters for the experimental reference data.

| $\theta_0$ | $I$ (nA) | $\alpha_1$ | $\alpha_2$ | $k_1$ | $k_2$ | $\omega$ |
|---|---|---|---|---|---|---|
| data#1 | 0.62±0.34 | 15.4 | -1.49 | 76 | 11.6 | 33.5 |
| data#2 | 0.16±0.32 | 7.9 | 1.24 | 190 | 38 | 14.6 |
| data#11 | 0.15±0.33 | 15.3 | -5.7 | 62 | 38 | 16.3 |

After the identification process, the estimated parameters were utilized in a MAT model to predict the spike train. Obtained results for comparing the reference spike train and predicted spike train are shown in Table (5) for the three data samples as shown above. Figures (10) and (11) display the experimental traces and predicted MAT model voltages while the similarity of the spike trains is pointed out by the spike times. A similar figure for data#1 is not provided because the neuron membrane potential data corresponding to that current has not been made available in the published dataset – only firing times are available.

Table 5: Comparison of the predicted spike train similarity and reference spike train.

|         | $N_{Data}$ | $N_{Model}$ | $N_{Coinc}$ | $\Gamma$ |
|---------|------------|-------------|-------------|----------|
| data#1  | 875        | 558         | 496         | 0.60     |
| data#2  | 268        | 246         | 230         | 0.88     |
| data#11 | 212        | 162         | 144         | 0.76     |

As discussed earlier in the paper, a set of intersection constraints was applied to the objective function per (18). To clarify the contribution of these intersection constraints, comparisons were made between spike trains predictions from models estimated with and without the use of constraints. Table (6) shows the comparative results using an experimental data sample. It can be seen that the use of constraints increases the accuracy of predicting spike times by more than 20%. In fact, the intersection constraints drive the estimator to avoid undesirable spikes and as a result the firing rate is decreased. Although unconstrained estimator predicts higher percentage of target spikes correctly, its performance, represented by coincidence factor, is worse than the constrained one.

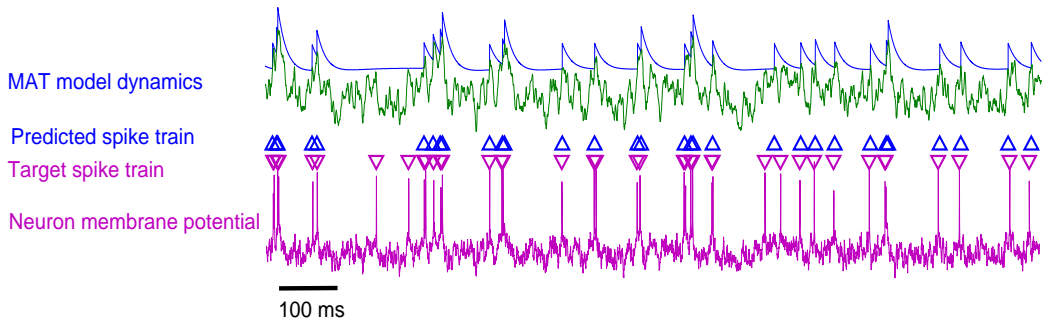

Figure 10: Model prediction for fluctuating current according to experimental data#2. The top row is the subthreshold (green) and threshold (blue) voltages trace of MAT model while predicted spike times are specified by blue triangles. The bottom row, indicates the experimental membrane potential (magenta) from a single neuron, while the actual spike train is marked by magenta triangles.

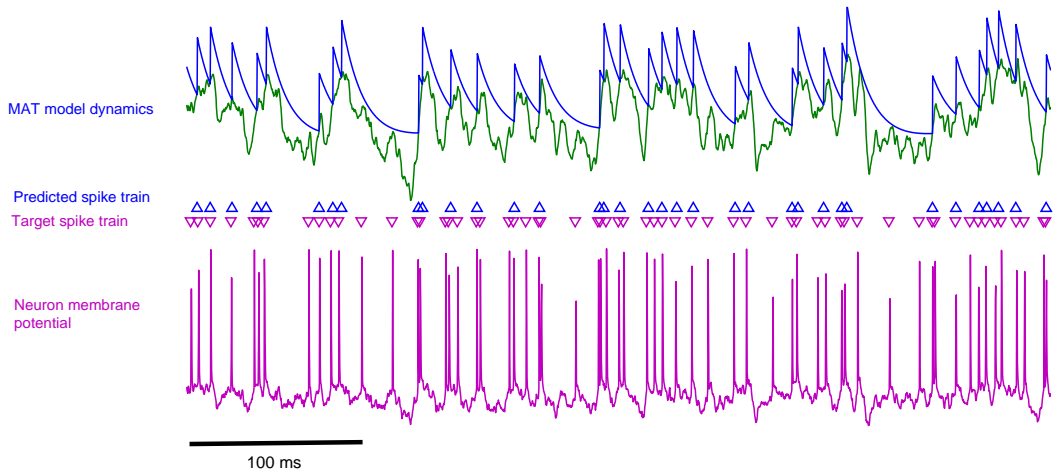

Figure 11: Model prediction for fluctuating current according to experimental data#11. The top row is the subthreshold (green) and threshold (blue) voltages trace of MAT model while predicted spike times are specified by blue triangles. The bottom row indicates the experimental membrane potential (magenta) from a single neuron, while the actual spike train is marked by magenta triangles.

Table 6: Comparison of predicted spike train for unconstrained/constrained estimation

|  | Unconstrained | Constrained |
|---|---|---|
| $N_{Coinc}/N_{Model}$ | 69% | 92% |
| $N_{Coinc}/N_{Data}$ | 88% | 51% |
| Spikes / sec | 176 | 77 |
| $\Gamma$ Coinc. Fac. | 0.35 | 0.59 |

## 5.4 Comparison with Existing Methods

In (Jolivet et al., 2008a), a benchmark test was established to facilitate a systematic comparison of methods and models in predicting the activity of rat cortical pyramidal neurons. The provided data set includes four different input currents which were generated based on (19). For each injected current, four trials were recorded to observe if the neuron fires with high reliability. To evaluate the quantitative predictive feature of our approach, we compare our proposed method performance with the benchmark test reported results. Figure (12) indicates the average performance of our method along with the benchmark test results on the whole data set. The raw $\bar{\Gamma}$ is computed by averaging the values of $\Gamma$ over the whole test set. Since for a certain input, the pyramidal neuron is more reliable than for the others (Mainen and Sejnowski, 1995), the normalized $\Gamma_A$ is also introduced. $\Gamma_A$ scales the raw $\Gamma$ according to reliability of the neuron which is evaluated with trial-to-trial variation of the neuron recordings (Kobayashi et al., 2009). Of the four submissions for the challenge, the auto regressive method (AR) demonstrated the best performance. The AR method uses a mathematical model to estimate the membrane potential of the neuron and then the spike times were predicted by ad-justing a dynamic threshold to the estimated membrane potential. The parameters of this model were determined so that the coincidence factor $\Gamma$ would be maximized. The carbon-copy method (CC) also yielded a good performance; however it does not benefit from any mathematical model. The CC method utilizes the mean and variance of fluctuating current and evokes a sequence of spike by considering the training data set. There were also two other anonymous submissions in the challenge whose performances are presented here. It is seen that the proposed method is superior to the challenge submissions by both the $\bar{\Gamma}$ and the $\Gamma_A$ metrics.

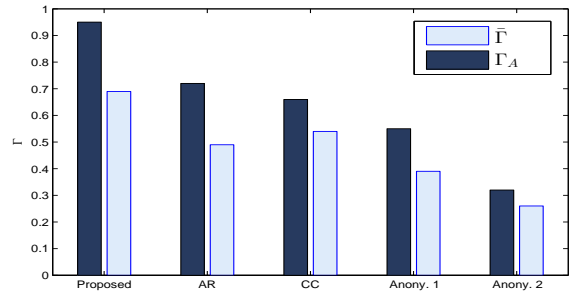

Figure 12: Comparison of the proposed method to results of the challenge.

## 5.5 Discussion

In this study, we took advantage of the MAT model which comprises two dynamics. Although the simplified subthreshold leaky integrator dynamics fail to consider many aspects of neuronal dynamics (Gerstner et al., 2014), the leaky integrator free parameters provide adequate strength to track the neuron membrane potential trace. Furthermore, the threshold dynamic of MAT model makes effective use of its multi-timescale feature. Biologically speaking, the multiple timescales can be regarded as surrogates for ionic currents such that different timescale values represent fast transient current, non-inactivating current, *etc.* (Yamauchi et al., 2011).

The proposed linear representation of the MAT model along with the novel objective function and constraints provides a framework for fitting the model to a single neuron recording in order to predict the quantitative features. The results for prediction of reference data generated from the exact MAT model confirm the validity of the manipulated equations and demonstrate the ability of the approach to find the best parameters even in the presence of noise. While the obtained results from experimental data demonstrate a high performance in predicting reference spike times, a detailed discussion is merited to analyze our approach more clearly. Our utilization of the objective function of (12) clarifies that the defined error function does not have an inherent mechanism to consider the non-spike moments. To overcome this issue, we added innovative constraints in our estimation procedure to control the threshold voltage from potential intersections with the subthreshold voltage during the period between two spikes. As shown above, the application of the constraints allows the identified model to generate the spike train with more confidence such that, by avoiding potential false positives, the firing rate is reduced which in turn leads to a higher predictive performance.

To evaluate of the proposed approach, we also compared the predictive performance of our identified model to the submissions for Quantitative Neuron Modeling. Our results show the superiority of the proposed linear method to predict reference spike trains. In addition to improvements in prediction, our method benefits from a convex cost function while at least some of the other submissions in the challenge utilize the non-convex $\Gamma$ Coincidence Factor as the cost function to maximize their prediction per-formance. Thus, our proposed approach also has a lower computational cost and a guaranteed global minimum which further underscores its superiority among the other methods. Finally, an obvious deficit of the MAT model is its inability to respond appropriately to rectangular and ramp currents. Therefore, future work will focus on model modification so it can response to not only fluctuating currents but also different input types.

# 6 Conclusions

In this paper, we proposed a constrained linear least squares algorithm to identify MAT model parameters, for predicting single neuron spike times. Our results show that the proposed identification method is robust to system noise and has the ability to find the best parameters to replicate the spike train. Moreover, the obtained experimental results indicates that our method has excellent performance in comparison to reported results from the Quantitative Neuron Modeling competition (Jolivet et al., 2008b). Convexity of the cost function is another advantage when compared with similar fitting approaches utilized currently by the neuronal community. While the MAT model succeeds in reproducing quantitative features of single neurons, it still lacks the capability to replicate different firing patterns; hence, there is room to investigate possible modifications of the model in the future.